\documentclass[twocolumn,preprint, ]{aastex63}  

\usepackage{amsmath,amstext}
\usepackage[T1]{fontenc}
\usepackage{apjfonts} 
\usepackage{graphicx}
\usepackage{natbib}
\usepackage{longtable}

\usepackage{microtype,longtable,verbatim,float,booktabs,graphicx}

\newcommand{\Ha}{\hbox{{\rm H}$\alpha$}}
\newcommand{\Hb}{\hbox{{\rm H}$\beta$}}
\newcommand{\Hg}{\hbox{{\rm H}$\gamma$}}
\newcommand{\Hd}{\hbox{{\rm H}$\delta$}}
\newcommand{\He}{\hbox{{\rm H}$\epsilon$}}

\newcommand{\HavNII}{\hbox{({\rm H}$\alpha$+{\rm [N}\kern 0.1em{\sc II}{\rm ]})}}

\newcommand{\HnFeV}{\hbox{{\rm H}9+{\rm [Fe}\kern 0.1em{\sc V}{\rm ]}}}

\newcommand{\HeH}{\hbox{{\rm H}8+{\rm He}\kern 0.1em{\sc I}{\rm }}}

\newcommand{\SII}{\hbox{[\ion{S}{2}]}}
\newcommand{\NII}{\hbox{[\ion{N}{2}]}}
\newcommand{\OI}{\hbox{[\ion{O}{1}]}}
\newcommand{\OII}{\hbox{[\ion{O}{2}]}}

\newcommand{\OIII}{\hbox{[\ion{O}{3}]}}
\newcommand{\OIIIa}{\hbox{[\ion{O}{3}]}\lambda4363}
\newcommand{\NeIII}{\hbox{[\ion{Ne}{3}]}}

\newcommand{\CIII}{\hbox{\ion{C}{3}]}}

\newcommand{\CIV}{\hbox{\ion{C}{4}}}
\newcommand{\NeIV}{\hbox{[\ion{Ne}{4}]}}
\newcommand{\NeV}{\hbox{[\ion{Ne}{5}]}}
\newcommand{\HeII}{\hbox{{\rm He}\kern 0.1em{\sc II}}}

\newcommand{\MgII}{\hbox{\ion{Mg}{2}]}}

\newcommand{\HeI}{\hbox{{\rm He}\kern 0.1em{\sc I}}}

\newcommand{\HII}{\hbox{{\rm H}\kern 0.1em{\sc II}}}

\newcommand{\NIV}{\hbox{{\rm N}\kern 0.1em{\sc IV}}}

\newcommand{\OIIIHb}{\OIII \lambda5007/\Hb}
\newcommand{\OIIIHg}{\OIII \lambda4363/\Hg}
\newcommand{\NeOII}{\NeIII/\OII}

\newcommand{\CIIIHe}{\CIII/\HeII}

\newcommand{\Mbh}{M_{\rm BH}}
\newcommand{\Msol}{M_{\rm \odot}}

\newcommand{\limit}[1]{\textcolor{gray}{#1}}

\begin{document}

\title{\large \bf Emission-Line Diagnostics at $z>4$: \OIII$\lambda$4363/\Hg}

\author[0000-0001-8534-7502]{Bren E. Backhaus}
\affil{Department of Physics and Astronomy, University of Kansas, Lawrence, KS 66045, USA}
\affil{Department of Physics, 196A Auditorium Road, Unit 3046, University of Connecticut, Storrs, CT 06269, USA}

\author[0000-0001-7151-009X]{Nikko J. Cleri}
\affiliation{Department of Astronomy and Astrophysics, The Pennsylvania State University, University Park, PA 16802, USA}
\affiliation{Institute for Computational and Data Sciences, The Pennsylvania State University, University Park, PA 16802, USA}
\affiliation{Institute for Gravitation and the Cosmos, The Pennsylvania State University, University Park, PA 16802, USA}

\author[0000-0002-1410-0470]{Jonathan R. Trump}
\affil{Department of Physics, 196A Auditorium Road, Unit 3046, University of Connecticut, Storrs, CT 06269, USA}

\author[0000-0002-5537-8110]{Allison Kirkpatrick}
\affiliation{Department of Physics and Astronomy, University of Kansas, Lawrence, KS 66045, USA}

\author[0000-0002-6386-7299]{Raymond C. Simons}
\affil{Department of Physics, 196A Auditorium Road, Unit 3046, University of Connecticut, Storrs, CT 06269, USA}

\author[0000-0002-7959-8783]{Pablo Arrabal Haro}
\affiliation{NSF's National Optical-Infrared Astronomy Research Laboratory, 950 N. Cherry Ave., Tucson, AZ 85719, USA}

\author[0000-0002-9921-9218]{Micaela B. Bagley}
\affiliation{Department of Astronomy, The University of Texas at Austin, Austin, TX, USA}

\author[0000-0001-5384-3616]{Madisyn Brooks}
\affil{Department of Physics, 196A Auditorium Road, Unit 3046, University of Connecticut, Storrs, CT 06269, USA}

\author[0000-0003-2536-1614]{Antonello Calabr\`o}
\affiliation{INAF Osservatorio Astronomico di Roma, Via Frascati 33, 00078 Monteporzio Catone, Rome, Italy}

\author[0000-0001-8047-8351]{Kelcey Davis}
\affiliation{Department of Physics, 196A Auditorium Road, Unit 3046, University of Connecticut, Storrs, CT 06269, USA}

\author[0000-0001-5414-5131]{Mark Dickinson}
\affiliation{NSF's National Optical-Infrared Astronomy Research Laboratory, 950 N. Cherry Ave., Tucson, AZ 85719, USA}

\author[0000-0001-8519-1130]{Steven L. Finkelstein}
\affiliation{Department of Astronomy, The University of Texas at Austin, Austin, TX, USA}

\author[0000-0002-3301-3321]{Michaela Hirschmann}
\affiliation{Institute of Physics, Laboratory of Galaxy Evolution, Ecole Polytechnique Fédérale de Lausanne (EPFL), Observatoire de Sauverny, 1290 Versoix, Switzerland}

\author[0000-0001-9187-3605]{Jeyhan S. Kartaltepe}
\affiliation{Laboratory for Multiwavelength Astrophysics, School of Physics and Astronomy, Rochester Institute of Technology, 84 Lomb Memorial Drive, Rochester, NY 14623, USA}

\author[0000-0002-6610-2048]{Anton M. Koekemoer}
\affiliation{Space Telescope Science Institute, 3700 San Martin Dr., Baltimore, MD 21218, USA}

\author[0000-0003-1354-4296]{Mario Llerena}
\affiliation{INAF - Osservatorio Astronomico di Roma, via di Frascati 33, 00078 Monte Porzio Catone, Italy}

\author[0000-0001-9879-7780]{Fabio Pacucci}
\affiliation{Center for Astrophysics $\vert$ Harvard \& Smithsonian, 60 Garden St, Cambridge, MA 02138, USA}
\affiliation{Black Hole Initiative, Harvard University, 20 Garden St, Cambridge, MA 02138, USA}

\author[0000-0003-3382-5941]{Nor Pirzkal}
\affiliation{ESA/AURA Space Telescope Science Institute}

\author[0000-0001-7503-8482]{Casey Papovich}
\affiliation{Department of Physics and Astronomy, Texas A\&M University, College Station, TX, 77843-4242 USA}
\affiliation{George P.\ and Cynthia Woods Mitchell Institute for Fundamental Physics and Astronomy, Texas A\&M University, College Station, TX, 77843-4242 USA}

\author[0000-0003-3903-6935]{Stephen M.~Wilkins} %
\affiliation{Astronomy Centre, University of Sussex, Falmer, Brighton BN1 9QH, UK}
\affiliation{Institute of Space Sciences and Astronomy, University of Malta, Msida MSD 2080, Malta}

\begin{abstract}

We use JWST Near-Infrared Spectrograph (NIRSpec) observations from the the Cosmic Evolution Early Release survey (CEERS), GLASS-JWST ERS (GLASS), and JWST Advanced Deep Extragalactic Survey (JADES) to measure rest-frame optical 
emission-line ratios of 90 galaxies at $z>4$.
The stacked spectra of galaxies with and without a broad-line feature reveal a difference in the $\OIII \lambda$ 4363 and $\Hg$ ratios. This motivated our investigation of the  $\OIIIHg$ vs $\NeOII$ diagram. We define two AGN/SF classification lines based on 1869 SDSS galaxies at $z\sim0$. After applying a redshift correction to the AGN/SF lines we find 76.8\% of BLAGN continue to land in the AGN region of the diagnostic largely due to the \NeOII\ ratio. However, 40.2\% of non-BLAGN land in the AGN region as well, this could be due to star forming galaxies having harder ionization of there are narrow line AGN which are not accounted for. This indicates the potential of the \NeOII ratio to continue classifying galaxies to $z\sim6$.
We further inspect galaxies without broad emission lines in each region of $\OIIIHg$ vs $\NeOII$ diagram and found that they have slightly stronger $\CIII\lambda$1908 fluxes and equivalent width when landing in the BLAGN region. However, the cause of this higher ionization is unclear. Additionally, we find that BLAGN are characterized by a higher ionization (at constant electron temperature) compared to non-broad line galaxies. 

\end{abstract}

  
\section{Introduction}\label{Intro}

Emission-line spectroscopy reveals a wealth of information about the formation and physical conditions of galaxies, such as the metallicity, ionization, and pressure of the interstellar medium (ISM). Emission lines can be used to derive other properties such as a galaxy's star formation rate (SFR)  \citep{Brin04,kenn12} and dust attenuation \citep{card89,Calz1999,Reddy2016,Shap2023Balm}. One classic way to analyze emission-line data is by comparing ratios of lines which are close together in wavelength, which makes the ratio less sensitive to dust attenuation. The most well-known emission-line diagrams are the BPT \citep{bald81} and VO87 \citep{veil87}, which compare $\OIII\lambda5007/\Hb$ with $\NII\lambda6583/\Ha$ or $\SII\lambda(6716+6730)/\Ha$, respectively, to diagnose the source of ionizing photons in z$<$2 galaxies. These diagrams tend to use the strongest emission lines in rest-frame optical spectra \citep{bald81,veil87,kauf03,kewl06}.

As we move to higher redshifts the strong rest-optical lines which are typically used at z$\sim$0 move to the observed near-IR. 
With the growing amount of JWST data at $z>4$ we can evaluate and define diagnostics to reveal how these galaxies' properties are evolving. One area of interest is understanding the large population of potential AGN at these high redshift and completing the AGN sample. 
Works such as \cite{Maiolino2023, Harikane2023,Koce2022,Kocevski2023,Taylor2024,Lars2023} have noted several broad-line AGN (BLAGN) using NIRSpec spectra at $z>4$, and the presence of these broad lines make for easy classification of BLAGN \citep{Greene2024, Taylor2024}. However, selecting narrow-line AGN at these high redshifts is more difficult. Many works have shown that traditional AGN emission-line diagnostic diagrams like the BPT and VO87 are not reliable for the high-redshift Universe, $z>3$ \citep{back22,Calabro2023,Kewley2013,Feltre2016,Hirsch2023,Cleri2023b}. 
The ISM of z>4 galaxies are generally more metal poor and have a higher ionization parameter. 
This increase in ionization from young stars in star-forming galaxies (SFG) and more metal-poor AGN-ionized gas causes the two populations to overlap in the BPT and VO87 diagram \citep{Backhaus2024, Kocevski2023,Feltre2016,Hirsch2023}. 
However, even with the increasing ionization of SFG the detection of even higher ionization emission lines, such as  \NeIV$\lambda$2424 (63.45eV) and [Ne V]$\lambda$3426 (97.12eV), may still be a safe tracer for the presence of an AGN \citep{Cleri2023a,Cleri2023b,Backhaus2024,Scholtz2023}. However, these high-ionization emission lines are typically weak and are difficult to measure even in the deepest JWST observations.

In order to fully understand how the physical conditions of galaxies change over cosmic time, emission-line diagnostics must be well defined and understood. Though our traditional methods are uncertain, JWST covers a large wavelength range ($\sim$ 1-5$\mu m$) and the sensitivity to access many rest-UV and optical emission lines for high-z galaxies that were inaccessible with previous observations. This gives us an opportunity to study new emission lines and properties of galaxy spectra that may reveal potential for a new diagnostic.

A potential emission line available within this new coverage and sensitivity is $\OIIIa$, which has been detected in multiple JWST spectra. \cite{Brinchmann2023} first noted that high $\OIIIa$ emission at z$\sim$8 could indicate the presence of an AGN. The $\OIIIa$ tends to be the strongest auroral line in the spectra and has a higher-energy electron transition than the \OIII$\lambda$4959,5007 doublet.
This emission line was further explored in \cite{Mazzolari2024}, which used galaxies at multiple redshifts and found that the $\OIIIHg$ emission line ratio could provide a sufficient but not necessary condition to find AGN. 
\cite{Ubler2023} proposed that the enhancement of $\OIII$$\lambda$ 4363 could be due to the higher ISM temperatures caused by an AGN. 

The presence of high-ionization UV lines in high-redshift galaxies have been connected to AGN, Wolf-Rayet stars, and starburst \citep{Maiolino2024,Saxena2020}. We explore the presence of the UV emission lines \CIV$ \lambda$1548,1550, \HeII $\lambda$1640, \NIV$\lambda$1718, and $\CIII$ $\lambda$1908 in high redshift galaxies. These lines could help further untangle galaxy populations by revealing if their ratios are more in line with the narrow-line regions of AGN or star-forming galaxies.  \cite{VandenBerk2001} showed that the strong ionizing background produced by AGN creates \CIII\ emission. At lower redshifts \CIIIHe\ is one of the most useful ways to separate star formation from harder ionizing sources \citep{Kewl19b}. At lower redshifts a galaxy is considered star forming if  log(\CIIIHe)$>1$. Other emission lines can reveal if there is a population of Wolf-Rayet stars via \NIV$\lambda$1718 emission \citep{Hainich2014}, or a population of very massive stars (VMS) from high \HeII$\lambda$1640 equivalent widths. 

The \CIIIHe\ ratio, as well as other ratios using \CIV, were further explored at higher redshifts and found to be good indicators of different ionization sources up to $z\sim4$ \citep{Hirschmann2019}. Additionally, \cite{Lefevre2019} which studied 3899 SFG's at $2<z<4$ and found that the four strongest \CIII\ emitters were associated with BLAGN, and that AGN feedback acts on strong \CIII\ emitters. 

The growing sample of high-redshift galaxies observed by the JWST gives us the opportunity to study and understand the ISM conditions. \cite{Trump22}, \cite{Cleri2023b}, \cite{Backhaus2024}, and \cite{Brin22} found higher ionization in the ISM for early galaxies ($z>5$) compared to the local ($z\sim0$) galaxies.
These early galaxies at $z>5$ have been shown to have moderate to low metallicity \citep{Backhaus2024,Sand2023Ex}. 

In this work we use NIRSpec Multi-Shutter Assembly (MSA) spectroscopy from Cosmic Evolution Early Release survey (CEERS), GLASS, and JADES observations to rest-frame optical and ultraviolet (UV) emission-line evolution and galaxy properties of 171 galaxies at $z>4$ to explore emission-line diagnostics ability to classify sources of ionization. In Section \ref{Data/Sample} we describe our sample selection, and we also establish our comparison samples of galaxies at $z\sim0$. In Section \ref{Stacked} we outline the calculation of our stacked spectra. In Section \ref{BPT} we review previous emission line diagnostics used at $z<2$. In Section \ref{OHNOvg} we define and investigate the $\OIIIHg$ vs. $\NeIII \lambda$3868/$\OII \lambda$ 3728 diagram. In Section \ref{ISM} we look into the physical conditions such as metallicity and temperature of galaxies at $z>4$. In Section \ref{CIII_diag} we investigate the ability of the \CIV, \HeII, \CIII\ emission lines to find AGN at $z>4$. In Section \ref{ISM} we derive electron temperature and metallicity estimates. We summarize our results in Section \ref{Summary}. In this work, we assume a $\Lambda$ cold dark matter cosmology with $\Omega_{M}=0.3$, $\Omega_{\Lambda}=0.7$, and $H_{0}=70$ $km s^{-1}$ $Mpc^{-1}$ \citep{plan15}.

\section{Observational Data and Sample} \label{Data/Sample}

\subsection{JWST Sample}

Our $z>4$ galaxy sample comes from the NIRSpec multi-object spectroscopy JWST observations taken by three programs: ERS-1345 (CEERS, PI: Steven Finkelstein), ERS 1324 (GLASS, PI: Tommaso Treu), and JWST GTO 121 (JADES, PI: Daniel Eisenstein, Nora Luetzgendorf).

The CEERS and JADES data are taken directly from the collaboration websites using their reductions while GLASS data is taken from the DAWN JWST Archive (DJA) \footnote{\url{https://dawn-cph.github.io/dja/spectroscopy/nirspec/}} using their v1.0 reductions. 

Information on the reductions of the NIRSpec data from CEERS can be found in \citet[][]{Arrabal2023SC} and Arrabal Haro et al. \textit{in prep.} The CEERS observations have 14 groups in NRSIRS2 readout mode per visit with a total exposure time of 3107~s. The CEERS observations used version 1.8.5 of the STScI Calibration pipeline and the Calibration Reference Data System (CRDS) mapping 1061. In Stage 1 of the reductions, the parameters of the jump step were modified to gain an improved correction of the “snowball” events. Additionally, a correction was performed from the detector 1/f noise and subtract the dark current and bias. Additionally in Stage 3 of the reductions custom extraction apertures where were created by visually identifying high SNR continuum or emission lines in their targets.

The GLASS observations are described in \cite{Treu2022}. Their observations have a 17682~s exposure time in all three high-resolution NIRSpec filters (R$\sim$2700). The DAWN JWST Archive (DJA) reductions of GLASS are described in \cite{Heintz2023} and \cite{deGraaff2024}. These are reduced using grizli and a custom-made pipeline (msaExp v. 0.6.7) \cite{Brammer2022}. Using the Stage 2 output from the MAST JWST archive the pipeline performs standard wavelength, flat-field and photometric calibrations using the CRDS mapping 1027. They scale and match the optimally-extracted 1D spectra to the available JWST/NIRCam photometry using a wavelength-dependent polynomial function. The pipeline additionally corrects the 1/f noise and bias levels of the exposures. The resulting 1D spectra are extracted using an inverse-weighted sum of the 2D spectra in the dispersal directions. Using a wavelength-dependent polynomial function the flux densities of the 1D spectra are scaled to match the integrated flux in the JWST/NIRCam passband for each filter. This improves the absolute flux calibration of the spectra and takes into account any potential slit-losses.

The JADES NIRSpec reductions are outlined in \cite{DEugenio2024}. The JADES observations cover 0.6-5.3 $\mu$m with the median-resolution gratings (R=500–1500). JADES is made up of multiple sets of observations. The exposure times are available in \cite{DEugenio2024}.


 
The differences in exposure time and resolution affects the SNR of the spectra, leading to different detection limits for emission lines in each dataset. However, the overall emission-line ratios should be comparable between the samples as we can assume differences in depth and reduction strategy will affect both emission lines in a ratio similarly.
 
\begin{figure}[h!]
\centering
\epsscale{1.}
\plotone{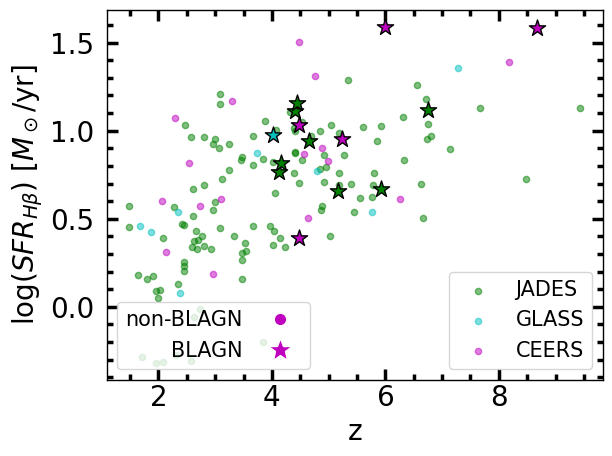}
\caption{Distribution of \Hb\ SFR and redshift for samples from JADES, CEERS, and GLASS NIRSpec observations. The galaxies hosting a BLAGN are denoted as a star. 
\label{fig:samp_dist}}
\end{figure} 

\begin{figure*}[tbp]
\centering
\epsscale{1.1}
\plotone{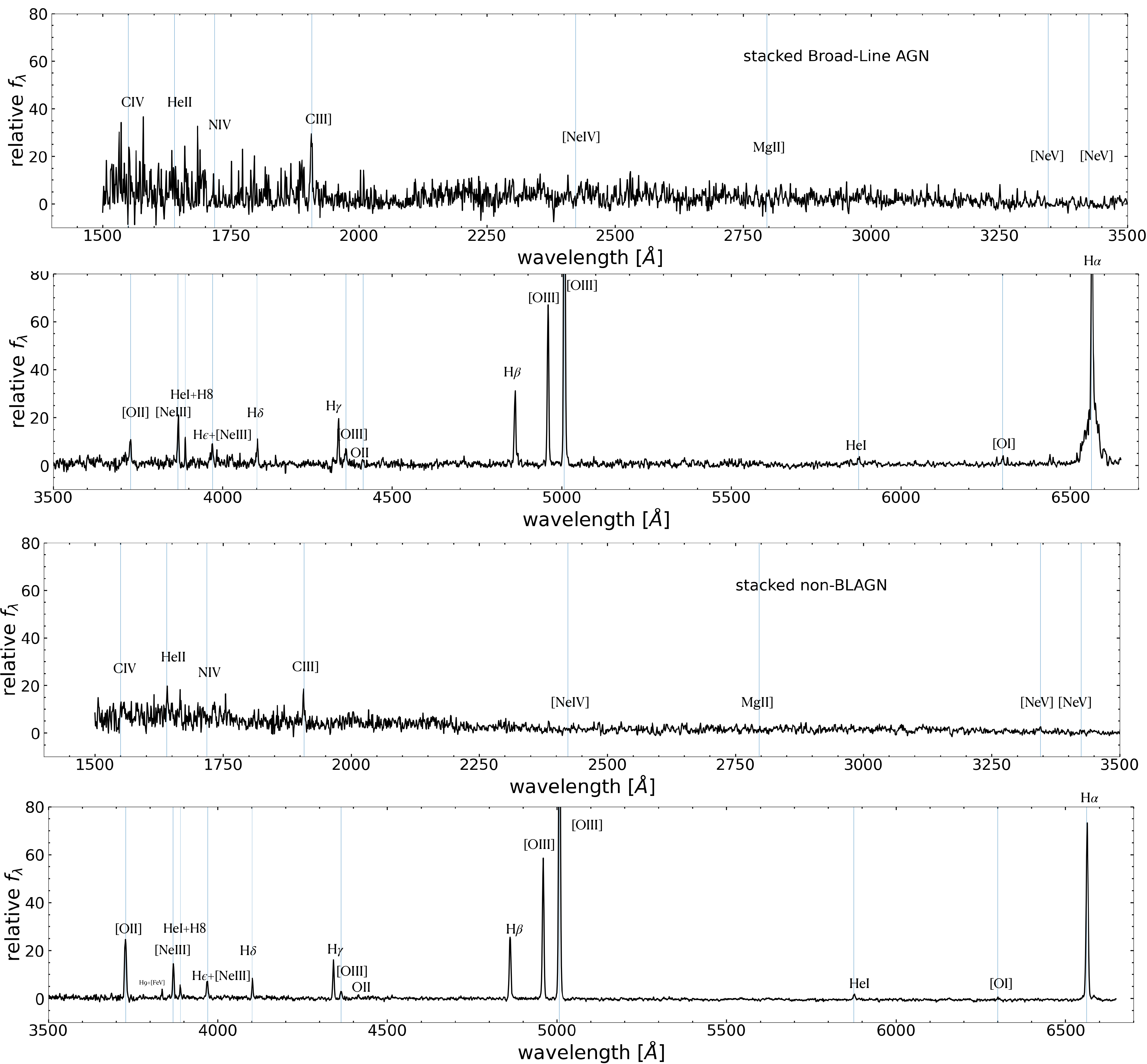}
\caption{Top: Stacked spectra for 9 BLAGN galaxies above $z>4$. The broad $\Ha$ feature can be seen in the bottom spectra, we also note even stacked a broad $\Hb$ component does not appear. Bottom: Stacked Spectra of 56 galaxies above $z>4$ without any broad-line features. 
\label{fig:stacked_BLAGN}} 
\end{figure*}

\subsubsection{Sample Selection}

Using the spectroscopic redshift catalogs of each sample CEERS \citep{Finkelstein2025}, JADES \citep{DEugenio2024}, GLASS (DJA) and apply a redshift constraint of $1.6<z<9$ as this covers the \OII, \NeIII, \Hg, and $\OIIIa$  emission lines in the G140M/H, G235M/H, and G395M/H gratings. We start with the initial redshift reported by the source of the data extractions, DAWN for GLASS, and the survey teams for CEERS and JADES. We then confirm or edit the spectroscopic redshift reported for each source used in our sample using the Gaussian centroid of the best fit of the brightest emission line in each spectrum, usually \OIII$\lambda$5008 or $\Ha$. We note for the GLASS sample which used the DAWN reductions and catalog that the redshift of 7 galaxies shifted by at least $\Delta{z} > 0.5$ from there reported redshift in the catalog, with 3 galaxies shifted $\Delta{z} > 4$.



We obtain secure redshifts and measure emission line fluxes using the best-fit Gaussian function (and associated uncertainties) using a Levenberg-Marquardt least-squares method implemented by the \texttt{SciPy curve fit} python code \citep{jones2001}. We require a SNR$>$3 for at least one emission line used in each ratio (typically $\Hg$ and $\OII$), allowing for upper or lower limits. In this sample 97.1\% of galaxies have SNR$>5$ for \OIII$\lambda$5007 or \Ha. This creates a sample of 177 (23 CEERS, 10 GLASS, 144 JADES) galaxies from $1.5<z<9.4$. The \Hb\ SFR and redshift of our entire sample is shown in Figure \ref{fig:samp_dist}.

\begin{figure*}[t!]
\centering
\epsscale{1.}
\plotone{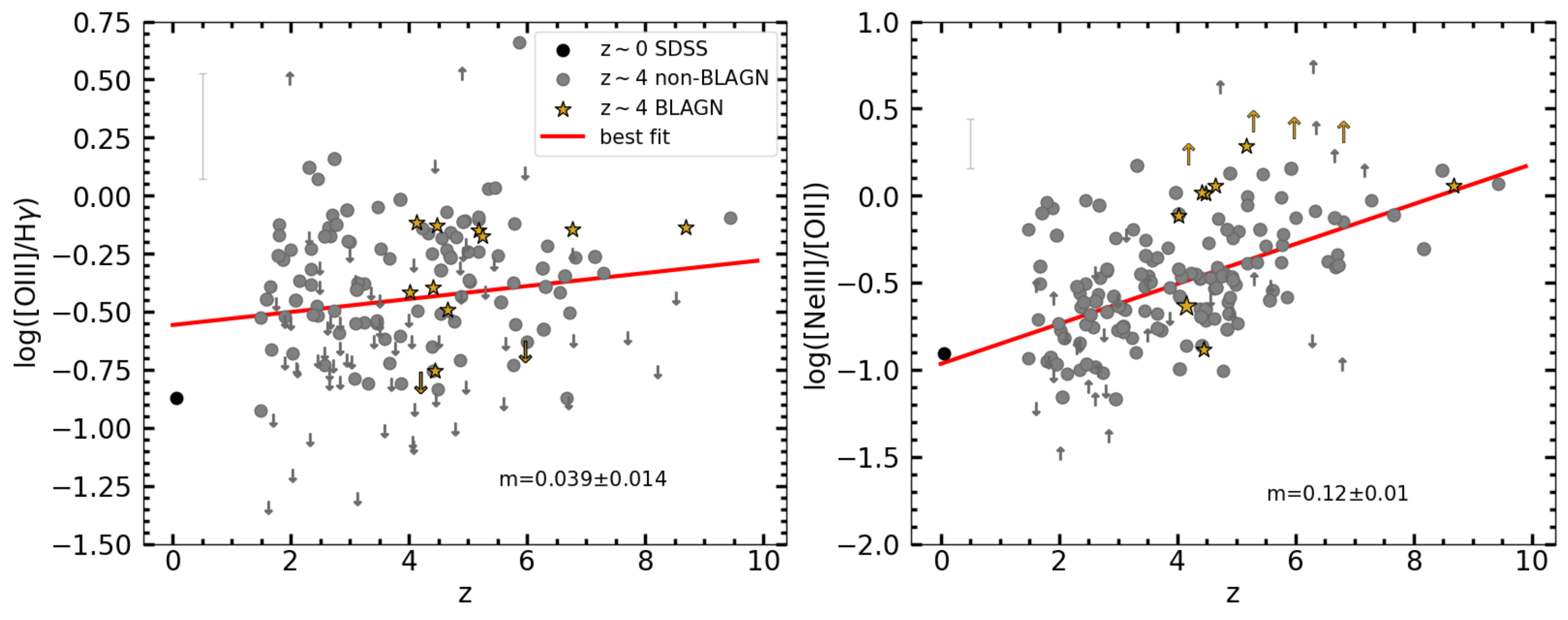}
\caption{Left: Evolution of the $\OIIIHg$ emission line over redshift. Right: Evolution of the \NeOII\ emission line over redshift. The arrows in these figures represent upper and lower limits in the emission line ratio. The red line represents the best fit is applied to the non-BLAGN galaxies (shown as gray points and arrows). The black point is the median ratio of the SDSS sample requiring SNR$>$3 in all emission lines. The average error bar for the emission line ratios is shown in the top left corner.
\label{fig:ELevo}} 
\end{figure*} 

For our high-redshift analysis we create a subsample of galaxies at $z>4$. The $z>4$ sample includes 90 (14 CEERS, 4 GLASS, 72 JADES) galaxies across all observational surveys. Our BLAGN sample starts with the BLAGN noted in \cite{Harikane2023} and \cite{Maiolino2023}. Both of these require a 5$\sigma$ detection of the broad-line component, in \cite{Harikane2023} they require the broad component to have a FWHM$_{Broad}$>1000km s$^{-1}$, while \cite{Maiolino2023} requires the broad component to be at least a factor of two broader than the narrow component. This list of potential AGN is then narrowed to galaxies that meet our emission line requirements. This leads to a sample of 13 galaxies shown as the stars in Figure \ref{fig:samp_dist}. We remeasure the \Ha\ FWHM of these 13 galaxies to confirm that they have a broad \Ha\ FWHM over 1000 km/s, and further inspection of the sample does not reveal any additional broad-line AGN.

Star formation rate (SFR) is calculated from the \Hb\ emission line, by following the \citet{kenn12} SFR relation for \Ha\ and $\Ha/\Hb=2.86$ (assuming Case B recombination, $T=10^4$~K, and $n_e=10^2$~cm$^{-3}$; \citealp{oste89}), resulting in the following equation:

\begin{equation} \label{Eq:SFRHb}
  \log({\rm SFR})[M_\odot/\mathrm{yr}] = \log[L(\Hb)] - 40.82\
\end{equation}

\noindent Due to the \Hb\ lines not being dust corrected, these SFR are lower limits.

With our previous cutoffs only 4 galaxies have multiple observations within one survey appeared in our sample. For the four galaxies, we chose the observation that had the best measured (highest SNR) emission lines. 

\subsection{SDSS $z \sim 0$ Sample}

We create a  $z\sim0$ comparison sample using the Sloan Digital Sky Survey \citep[SDSS;][]{york00} to cover the local Universe, z$\sim$ 0.05 \citep{Kauf03a,Kauf03b,Brin04}. This sample will be used to define the AGN/SF lines for our emission line diagnostics. The SDSS data set is observed using the 2.5 m telescope at Apache Point Observatory to cover 14,555 deg$^2$ in the sky with $R \sim 2000$ over $3800<\lambda<9200$ \AA\ \citep{smee13}.

Our SDSS dataset uses the emission-line measurements and redshifts that are computed by \cite{bolt12}. These are computed using a stellar template to correct the continuum for stellar absorption.

Our SDSS sample requires SNR$>3$ in $\Ha$, \SII, \OIII$\lambda$5007, and $\Hb$ so they can be separated as AGN or SF in the \citep{veil87} [VO87] diagnostic, $\OIIIHb$ vs $\SII$/$\Ha$. Additionally,  a SNR$>3$ for $\OIIIa$, $\Hg$, $\NeIII$, and $\OII$  is required, resulting in our z$\sim$0 sample having 1,869 galaxies. 


\subsection{Stacked Spectra}\label{Stacked}

We create stacked spectra galaxies showing those with a broad-line feature (Top panel of Figure \ref{fig:stacked_BLAGN}) and those without a broad-line feature (Bottom pabel Figure \ref{fig:stacked_BLAGN}). In order to be included in the stack, the individual galaxies are required to have wavelength coverage from 1500 to 6650 \AA\ and have the $\OIII\lambda$5007 emission line. These spectra are normalized to have the same \OIII$\lambda$5007 flux, and the stack is constructed from the median of the spectra to be more robust to outliers.



\section{Redshift Evolution in Emission-Line Diagnostics} \label{BPT}

At higher redshifts, previous emission-line diagnostics need to be revisited. Classical diagrams like BPT and VO87 face issues of less galaxies possessing strong enough $\SII\lambda$6716+6730 and $\NII\lambda$6583 emission due to lower gas phase metallicity at high redshift, $z>4$. 
There is another issue resulting in SFG having an increased ionization at higher redshifts \citep{Kewley2015,Sand2023Ex,Kaasinen2018}. In \cite{back22}, the unresolved VO87 ($\OIIIHb$ vs $\SII$/($\Ha$+$\NII$)) diagram was shown to no longer be able to separate ionization sources at $z\sim1.5$. This is due to galaxy ISMs having harder ionization and lower metallicity at higher redshifts. 

Figure \ref{fig:ELevo} plots $\OIIIHg$ (left) and $\NeOII$ (right) against redshift. The red line is a linear regression fit line to the NIRSpec galaxies without a BLAGN feature using \texttt{scipy} \citep{jones2001}. These fit lines indicate a 0.16 dex increase in $\OIIIHg$ and a 0.50 dex increase of $\NeOII$ between z$\sim0$ to z$\sim$4.

\section{$\OIIIHg$ vs $\NeOII$ Diagram} \label{OHNOvg}

With previously defined diagnostics struggling to separate SFGs from AGN sources, we evaluate a new diagnostic. To emphasize the differences between galaxies with and without an AGN we compare the stacked spectra of 9 known BLAGN galaxies to 56 galaxies with no broad line feature, as shown in Figure \ref{fig:stacked_BLAGN}. 

Comparing the two figures shows that the ratio between the $\NeIII$ and $\OII$ emission lines differ, with the BLAGN stack having a $\log(\NeOII)$ of 0.21 and the non-broad line stack with -0.41.
This implies that \NeOII\ remains a useful diagnostic for distinguishing AGN from star-forming galaxies. Another ratio of interest is \OIII\ $\lambda$4363 and \Hg; $\OIII$4363/5007 can be used to estimate the electron temperature of the ISM.

In \cite{Mazzolari2024} they test multiple emission line diagnostics including $\OIIIHg$ vs $\NeOII$. They used SF photoionization models from \cite{Gutkin2016} and narrow-line AGN models from \cite{Feltre2016} to show that high values of $\OIIIHg$ is a sufficient condition to identify the presence of AGN. However, the photoionization models indicate that low values of $\OIIIHg$ can be produced by both SF galaxies and narrow-line AGN.

\begin{figure}[tbp]
\centering
\epsscale{1.1}
\plotone{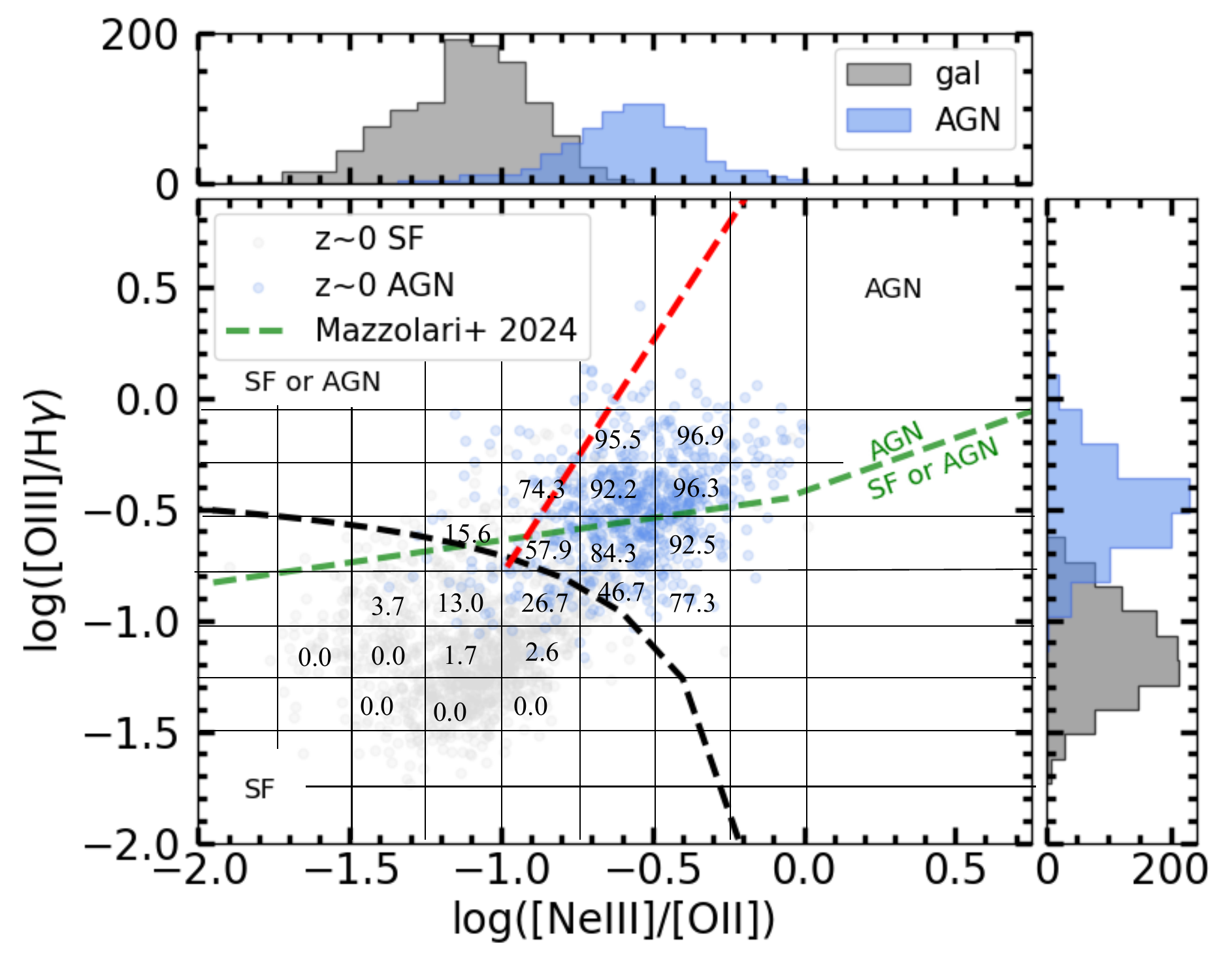}
\caption{$\OIIIHg$ vs. $\NeOII$ diagram of z$\sim$0 galaxies with a SNR$>$3 for all emission lines. Numbers within the grid show the percent chance of a being an AGN (blue point). A galaxy is marked as an AGN if it resides in the AGN region of the VO87 diagram. Galaxies noted as AGN in the VO87 diagram have higher $\OIIIHg$ and $\NeOII$. The green dashed line is the AGN/SF line defined in \cite{Mazzolari2024}.
\label{fig:OHNOvg_z0_final}} 
\end{figure}

The emission-line diagnostic we explore is $\OIIIHg$ vs $\NeOII$, for galaxies in the local universe $z\sim0.05$. Figure \ref{fig:OHNOvg_z0_final} plots 1869 galaxies that have S/N$>3$ for the respective emission lines. 
Figure \ref{fig:OHNOvg_z0_final} reveals the parameter space of $z\sim0$ SF galaxies and AGN, and has a similar distribution to \cite{Mazzolari2024} which used SDSS data using SNR$>5$. 
In Figure \ref{fig:OHNOvg_z0_final}, we indicate AGN identified by the VO87 diagram, which is shown to work at $z\sim0$ \citep{veil87}. 

Figure \ref{fig:OHNOvg_z0_final} shows that VO87-classified AGN prefer higher $\OIIIHg$ and $\NeOII$. We run an Anderson-Darling test to further test the $\OIIIHg$ vs $\NeOII$ diagrams ability to classify AGN from non-AGN. The Anderson-Darling test returns a p-value of 0.001 for both emission-line ratios. This indicates that, at least at lower redshifts, AGN and SFR are not from the same parent distribution and the $\OIIIHg$ vs $\NeOII$  diagram can separate AGN from SF galaxies. This is further shown in the histograms in the diagram. Additionally, we bin the SDSS sample and calculate the number of AGN in each bin. These numbers are shown in Figure \ref{fig:OHNOvg_z0_final}, and find a general increase in AGN percentage as we move to higher $\OIIIHg$ and $\NeOII$. 

From this diagram we define two lines. We first define a line that encompasses star-forming galaxies with less than 5\% AGN contamination, shown in black in Figure \ref{fig:OHNOvg_z0_final} and expressed as:

\begin{equation} \label{Eq:OHNOvg_line1}
  \log\left(\frac{\OIIIa}{\Hg}\right)=\frac{0.9}{1.99\log\left(\frac{\NeIII}{\OII}\right)-0.12}+0.28
\end{equation}

We create a second line that separates regions of 50\% AGN and SF galaxies. This line is shown in red and is defined as:

\begin{equation} \label{Eq:OHNOvg_line2}
  \log\left(\frac{\OIII}{\Hg}\right)=2.11\log\left(\frac{\NeIII}{\OII}\right)+1.308
\end{equation}

Galaxies that lie above the black line (Equation 2) and to the right of the red line (Equation 3) have 85\% of the AGN.

The green dashed line in Figure \ref{fig:OHNOvg_z0_final} is the AGN/SF line defined in \cite{Mazzolari2024}, which is based on the narrow-line AGN and SF photoionization models and observational sample distributions mentioned previously. Above this green line are only AGN photoionization models and below this line has a mixture of AGN and SF models. Additionally, this line ensures that the contamination of SDSS SF galaxies in the AGN region is less than 1-2\%. Comparing the \cite{Maiolino2024} AGN/SF line to our $z\sim0$ sample, we find that 62.1\% of AGN are in the AGN, whereas 37.9\% of AGN are in the AGN or SF region. 

\begin{figure}[t!]
\centering
\epsscale{1.1}
\plotone{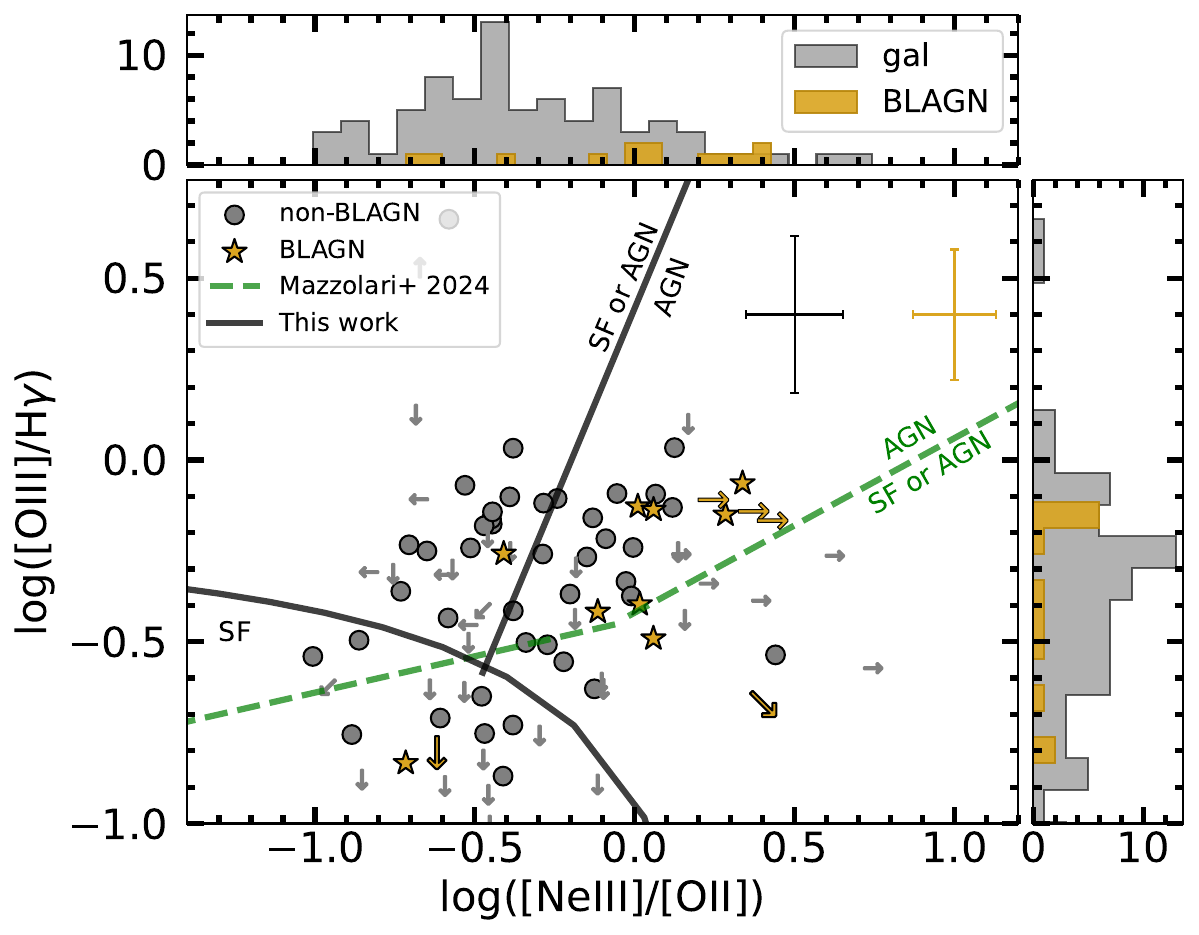}
\caption{$\OIIIHg$ vs \NeOII\ for $z>4$ galaxies. BLAGN are colored gold whereas other galaxies are gray. The average error bar for BLAGN and other galaxies are shown in the top right corner. The histogram on either axis shows the distribution between these two samples. We redshift correct the SF/AGN lines defined in Figure \ref{fig:OHNOvg_z0_final} using the evolution of the emission lines shown in Figure \ref{fig:ELevo}. High-redshift BLAGN appear to prefer higher $\NeOII$. The AGN region defined by Equations 2 and 3 for $z \sim 0$ galaxies contains most of the $z>4$ BLAGN.
\label{fig:OHNOvg}} 
\end{figure} 

Before looking at the $\OIIIHg$ vs $\NeOII$ diagram at $z>4$, we modify the AGN/SF lines defined at $z\sim0$ from Figure \ref{fig:OHNOvg_z0_final}. This modification to the AGN/SF line uses the empirical relation of the emission-line ratios and redshift shown in Figure \ref{fig:ELevo}. This modification considers the evolution of emission-line properties for non-AGN galaxies, associated with changes in ISM properties as a function of redshift.
We note non-broad-line galaxies may contain SF and NLAGN. To do this we find the difference between the z=0 and z=4 flux ratios according to the best linear fit. This change in emission line ratios are accounted for by the change in the x and y intercept accordingly. The green Mazzolari line is not shifted, in Figure \ref{fig:OHNOvg} as this line is based on models rather than observations.

With this the $\OIIIHg$ vs $\NeOII$ diagram at $z>4$ can be evaluated, using 89 galaxies observed by CEERS, GLASS, and JADES. Figure \ref{fig:OHNOvg} plots this sample and makes note if a galaxy has a broad-line feature indicating a BLAGN. The evolution-matched SF/AGN lines are shown as black lines. Additionally, Figure \ref{fig:OHNOvg} has the SF/AGN line defined in Table 2 of \cite{Mazzolari2024} as a green dashed line which considered both the photoionization models and the observational sample distributions.

Figure \ref{fig:OHNOvg} shows that the majority of our BLAGN land in the AGN region. In the AGN region of our $\OIIIHg$ vs $\NeOII$ diagnostic, 33.3\% of the galaxies are BLAGN where as there is 3.7\% on the left side. Running Anderson-Darling tests between the BLAGN and the rest of the sample for $\log{(\OIIIHg)}$ results in a p-value of 0.25 while for $\log{(\NeOII)}$ it finds a p-value of 0.001. High-redshift BLAGN still lie in the AGN region defined at $z \sim 0$ as 79\% of BLAGN land in this region, and the A-D test indicates that BLAGN have higher \NeOII\ ratios than SF galaxies. However, 40.2\% of galaxies in the canonical AGN region do not have broad emission lines. This indicates that the $\OIIIHg$ vs $\NeOII$ does not provide a robust separation of (broad-line) AGN and non-broad line galaxies at $z>4$.
The \NeOII\ ratio may be able to be used to note potential BLAGN sources if the \Ha\ line is not available.

When using the AGN/SF line noted in \cite{Mazzolari2024}, 71.4\% of the BLAGN land in the AGN region, while 21.6\% of the BLAGN are in the SF or AGN region. Additionally, 63.5\% of non-BLAGN land in the \cite{Mazzolari2024} AGN region.
It is worth noting our current sample does not account for any potential narrow-line AGN. This leaves two possibilities for these non-BLAGN galaxies in the AGN region. The first is that high redshifts the SF galaxies have higher ionization which makes the \NeOII\ and $\OIIIHg$ ratios confused with AGN. The other is that these non-BLAGN in the AGN region are narrow-line AGN. However, in order to distinguishing between these two scenarios further evidence is needed, such as eep spectroscopy of high-ionization AGN lines (e.g., \NeV$\lambda$3426).


\subsection{Photoionization Models}\label{Models}

We compare our sample to Cloudy photoionization models \citep[C23.01][]{Gunasekera2023}. The computations were performed using the Texas A\&M High Performance Research Computing cluster Grace\footnote{See \url{https://hprc.tamu.edu/kb/User-Guides/Grace}, where the namesake of the cluster is  \href{https://en.wikipedia.org/wiki/Grace_Hopper}{Grace Hopper}.}. The full description and public release of these models is in a forthcoming work (Cleri et al. in preparation).

All models are computed using \cite{Grevesse2010} solar abundance ratios. All models assume a gas density of $n_\mathrm{H} = 10^2 ~\mathrm{cm}^{-3}$. The AGN models are computed using ionizing SEDs of model AGN of varying black hole masses \citep{Done2012,Cann2018}. The SEDs can be retrieved from \texttt{XSPEC} \citep{Arnaud1996} using the \textsc{optxagnf} command, and are computed over the following nebular parameters:

\begin{itemize}
    \item black hole mass: $\log \Mbh/\Msol=[5, 6, 7, 8]$
    \vspace{-2mm}
    \item ionization parameter: $\log(U) = [-3, -2.5, -2, -1.5]$ 
    \vspace{-2mm}
    \item metallicity: $Z/Z_\odot = [0.05, 0.1, 0.2, 0.3, 0.4, 0.5]$
\end{itemize}

The stellar models are computed using ionizing SEDs from BPASS v2.2.1 \cite{Stanway2018}. We choose the single burst stellar populations with a \cite{salp55} initial mass function ($\alpha_{2}=-2.35$) and high-mass cutoff $M_{max} = 300 M_{solar}$, computed over the following parameters:  
\begin{itemize}
    \item stellar age: $\log$(age/years)=[6, 6.2,  6.4,6.6, 6.8]
    \vspace{-2mm}
    \item ionization parameter: $\log(U) = [-3, -2.5, -2]$
    \vspace{-2mm}
    \item metallicity: $Z/Z_\odot=[0.05, 0.1, 0.2, 0.3, 0.4, 0.5]$
\end{itemize}
where the stellar and gas-phase metallicities are kept equal. 

\begin{figure*}[t!]
\centering
\epsscale{1.1}
\plotone{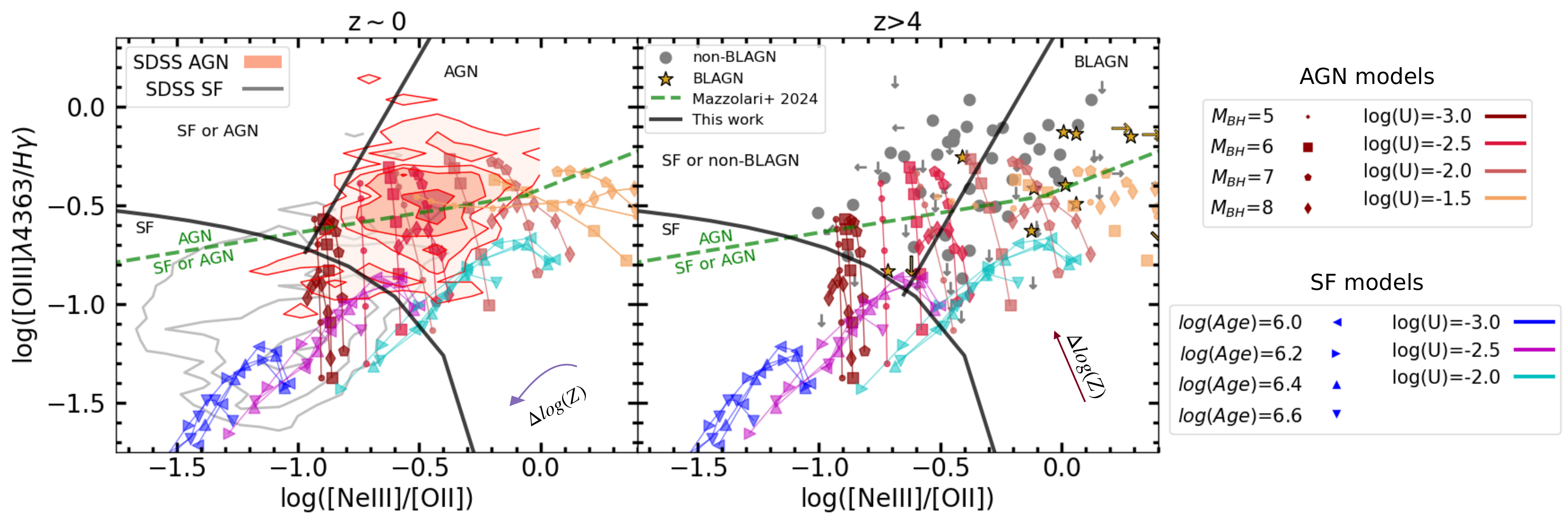}
\caption{$\OIIIHg$ vs $\NeOII$ with AGN and SF photoionizaion models from Cleri in prep. The inset vectors in the lower right corner indicate the direction of increasing  in metallicity for the SF models in the left panel and AGN models in the right panel. Left: $z\sim0$ SDSS galaxies with SNR>3 in all emission lines, where gray contours represent SF galaxies and red contours represent AGN. The SF galaxies are best represented by SF models with low to moderate ionization and low to moderate metallicity. Meanwhile the AGN are best represented by AGN models with moderate ionization and $0.2<Z/Z_\odot<0.5$. Right: $z>4$ galaxies are best represented by AGN models. The BLAGN are best represented by AGN models with high ionization and moderate to high metallicity, while galaxies without a broad line feature are best represented by AGN models with low to moderate ionization and moderate to high metallicity. 
\label{fig:Cleri_mod}} 
\end{figure*}

\begin{figure*}[tbp]
\centering
\epsscale{1}
\plotone{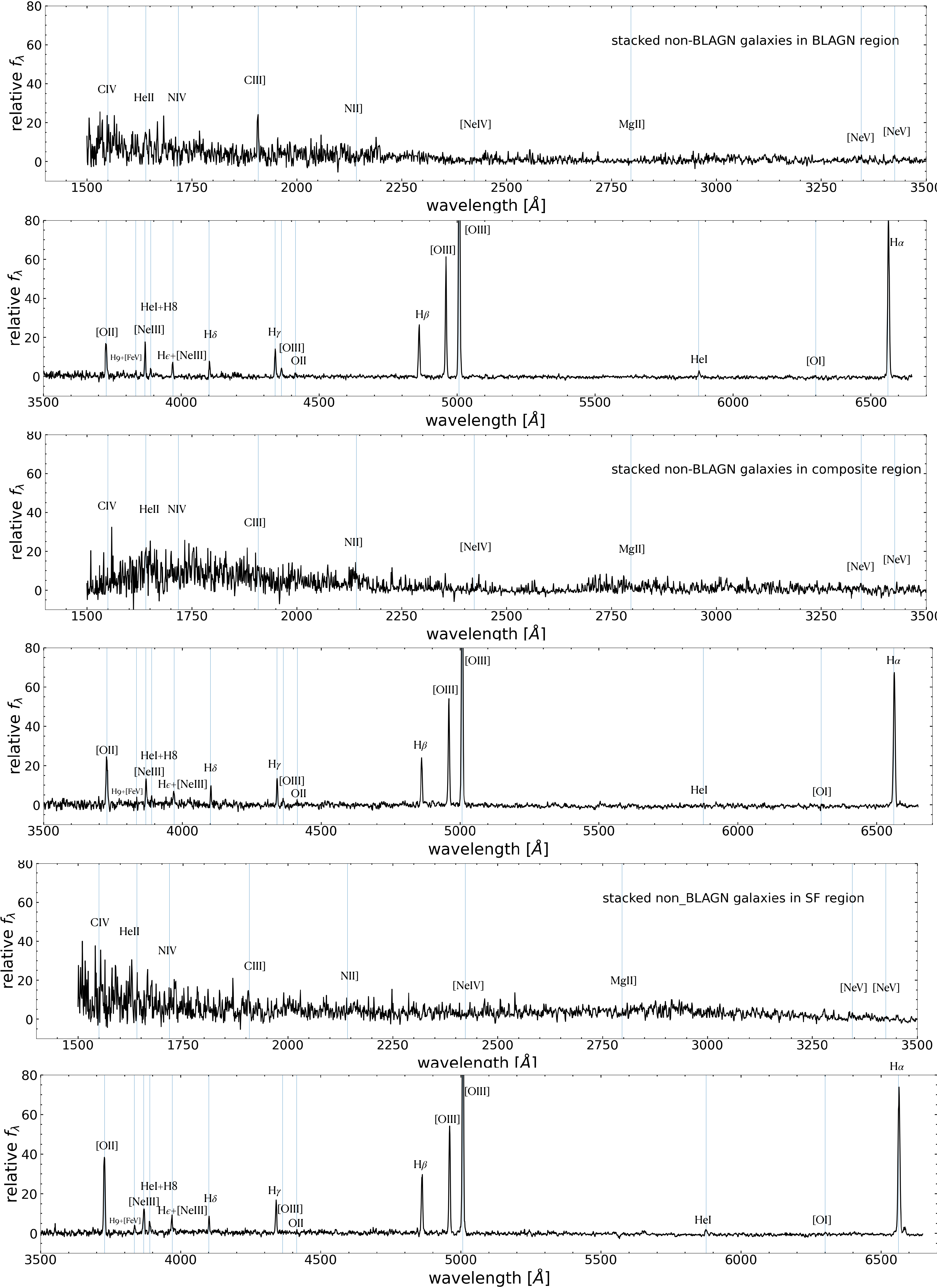}
\caption{Stacked spectra of galaxies that do not have a broad emission line feature based on region of the $\OIIIHg$ vs $\NeOII$ diagram they occupy. Top: 18 stacked galaxies in the BLAGN region of the $\OIIIHg$ vs $\NeOII$ diagram. Middle: 19 stacked galaxies in the composite region of the $\OIIIHg$ vs $\NeOII$ diagram. Bottom: 18 stacked galaxies in the SF region of the $\OIIIHg$ vs $\NeOII$ diagram. The emission line flux for these stacked spectra are shown in Table \ref{tab:stacked_lines}.
\label{fig:stacked_spectra}} 
\end{figure*} 


The left hand panel of Figure \ref{fig:Cleri_mod} compares the $z\sim0$ $\OIIIHg$ vs $\NeOII$ diagram to AGN and SF photoionization models from Cleri in prep. and shows that the SF region is best described by SF models with low to moderate ionization $-3<log(U)<-2.5$ and low to moderate metallicity $0.1<Z/Z_\odot<0.2$. Meanwhile the $z\sim0$ AGN region of the $\OIIIHg$ vs $\NeOII$ is best described by the AGN models with moderate ionization $log(U)=-2.5$ and moderate metallicity $0.1<Z/Z_\odot<0.3$.

The right hand panel of Figure \ref{fig:Cleri_mod} compares the $z>4$ $\OIIIHg$ vs $\NeOII$ diagram to the same AGN and SF photoionization models and shows that the AGN models also land in the composite region of the $z>4$ $\OIIIHg$ vs $\NeOII$ diagram. The composite region of the $z>4$ $\OIIIHg$ vs $\NeOII$ diagram is best described by AGN models with moderate ionization parameter $-2.5<log(U)<-2$ and a metalicity  parameter of $0.2<Z/Z_\odot<0.5$. In contrast, the $z>4$ BLAGN region of the $\OIIIHg$ vs $\NeOII$ diagram is best modeled by the AGN models with high ionization $-2<log(U)<-1.5$ and moderate to high metallically $0.2<Z/Z_\odot<0.5$.

The AGN/SF lines we define in this work do not separate the AGN and SF photoionization models: both the AGN and SF regions of the $\OIIIHg$ vs $\NeOII$ diagram contain both SF and AGN models within them. By comparison, the AGN/SF line from \cite{Mazzolari2024}, which was defined based on primarily photoionization models, does a much better job at separating the AGN and SF models from Cleri in prep.
 
These models reveal that galaxies at increasing redshifts are better described by models with increased ionization. 
More specifically, AGN photoionization models show that high-redshift galaxies regardless of whether the galaxy possesses features indicating an AGN, such as broad hydrogen lines or higher ionization lines, are better fit by harder ionizing sources. 
In addition to or instead of increased AGN activity, it is possible that $z>4$ galaxies have increased ionization from other sources, including massive low-metallicity stars, shocks, and/or higher ISM density \citep{Papo2022,Sand2023Ex,Kewl19}.

We relabel the three regions of the $\OIIIHg$ vs $\NeOII$ diagram to represent the $z>4$ galaxy observations and photoionization models. BLAGN have a preference for higher levels of \NeOII\ and $\OIIIHg$, and there is a preference for the SF models to have lower $\OIIIHg$ and lower to moderate $\NeOII$. We explore the non-broad line galaxies in each of these regions further to try to narrow down potential ionization sources in the next Section.

\subsection{Ultraviolet Emission Lines} \label{CIII_diag}


$\CIII$ and \HeII\ have been investigated in several works as a potential emission-line diagnostic \citep{Lefevre2019, Feltre2017}. A potential diagnostic such as $\CIII/\CIV$ vs $\CIII/\HeII$ would probe different regions of ionization as the ionization energies of \CIV, \HeII, and \CIII\ are 64.5eV, 54.4eV, 47.9eV respectively. Due to most individual galaxies not having a 1$\sigma$ measurement in two of the three lines for this diagnostic, we instead focus on the stacked spectra of our sample. 

Looking back at Figure \ref{fig:stacked_BLAGN}, the UV lines \CIII$\lambda$1909, \HeII$\lambda$ 1640, and \CIV$\lambda$1550 in the BLAGN stacked spectra also appeared to be stronger. The BLAGN stack has a \CIII\ emission that is 12.2\% the strength of the \OIII$\lambda$5007 flux and a rest-frame \CIII\ equivalent width of 15.9 \AA, while the non-BLAGN have a \CIII\ emission flux that is 1.2\% of the strength of \OIII$\lambda$5007 with a rest-frame \CIII\ equivalent width of 7.3 \AA.
This indicates that BLAGN at $z>4$ have a more significant \CIII\ flux and equivalent width than non-broad-line galaxies. This matches observations made in \cite{Lefevre2019} for AGN at $z\sim1-2$.

Figure \ref{fig:stacked_spectra} shows the difference between the stacked spectra of galaxies with no broad-line feature that reside in the AGN, composite, and star-forming region. Table \ref{tab:stacked_lines} lists the flux measurements relative to the \OIII$\lambda$5007 emission line for the stacked spectra shown in Figures \ref{fig:stacked_spectra} and \ref{fig:stacked_BLAGN}.

\begin{table}[h]
    \centering
    \begin{tabular}{c|C|c|c|c}
    & \textbf{SF} & \textbf{BLAGN}&  \textbf{Composite}  & \textbf{BLAGN}\\
    & \textbf{region} & \textbf{region}&  \textbf{region}  & \\
    \hline
     $\NII\lambda$6583 & 2.6 &  \limit{0.3} & 0.5 & \limit{--}\\
     
     $\Ha$ & 58.8 & 57.0 & 51.4 & 78.4\\
     
     $\OI\lambda$6300 & \limit{0.8} & \limit{0.9}  & \limit{0.5} & 1.7 \\
     
     $\HeI\lambda$5875  & 2.2 &1.9  & \limit{0.7}  & 2.1\\
     
     $\Hb$ & 19.2  & 16.4 & 14.7 & 15.4\\
     
     
     $\OIII\lambda$4363  & \limit{0.2} &2.8 & 2.8 & 4.2\\
     
     $\Hg$ & 8.9 &7.7 & 6.7 & 7.6\\
     
     $\Hd$ & 3.7 &3.5 & 3.9 & 5.1\\
     
     $\He$+$\NeIII\lambda$3967 & 4.5 &3.5 & 3.8 & 3.6 \\
     
    $\HeI$+H8 & 1.9 &0.9 & 0.8&  3.1 \\
    
    $\NeIII\lambda$3868 & 7.1 &7.4 & 6.1 & 8.0\\
    
    $\OII\lambda$3726,3728 & 27.0 &11.5 & 17.8 & 5.2 \\
    
    
    $\CIII$$\lambda$1909  & 3.9/4.5\AA & 8.0/13.3\AA & 2.3/3.9\AA & 12.2/15.9\AA \\
    
     $\HeII$$\lambda$1640 & \limit{0.3/0.9$\AA$} & 0.8/3.6\AA & 5.4/2.3\AA & 4.0/1.4\AA\\
     
     $\CIV$$\lambda$1550 & \limit{-/-} & \limit{1.5/4.5\AA} & \limit{--/--} & \limit{1.4/0.6\AA}\\
    \end{tabular}
    \caption{Emission-line flux relative to the \OIII$\lambda$5007 line, expressed as $100\times f(\mathrm{line})/f\OIII$, for stacked spectra of non-BLAGN galaxies which land in the BLAGN and SF regions of the $\OIIIHg$ vs $\NeOII$, and stacked measurement from BLAGN. The UV lines \CIII\, \HeII\, and \CIV\ also have the rest-frame equivalent width in $\AA$ after the flux.}
    \label{tab:stacked_lines}
\end{table}

The galaxies with no broad-line feature that land in the AGN region have stronger $\CIII$ emission. The measured rest-frame equivalent width of the \CIII\ line is 4.5 \AA\ for galaxies without a broad-line feature in the composite region and 13.3 \AA\ for galaxies without a broad line feature in the AGN region. As stated previously, a higher \CIII\ equivalent width is a feature seen in Figure \ref{fig:stacked_BLAGN}, and has been linked to AGN. 
This indicates that galaxies without a broad line feature in the BLAGN region may be narrow-line AGN. \cite{Lefevre2019} found that a \CIII\ rest-frame equivalent width above 10\AA\ can be a possible indicator of an AGN, as \CIII\ equivalent width correlates with the appearance of higher-ionization emission lines (such as $\CIV$). Non-BLAGN in the AGN region has the highest upper limit of $\CIV$, though no detection. No spectra show any indication of a \OIII$\lambda$1663 emission line either. 
To fully untangle the ionization sources, deeper observations on $z>4$ galaxies are needed to properly measure these UV lines. This would allow for individual detections that can lead to further diagnostics with a ratio using $\CIII$.

The presence of \CIII\ has not just been linked to AGN. Other studies have also shown that increasing \CIII\ equivalent width could be caused by younger massive metal-poor stars \citep{Schaerer2003,Ravindranath2020}. This can be further supported by non-broad line galaxies in the AGN region of the $\OIIIHg$ vs $\NeOII$ diagram having less \NII$\lambda$6583 emission compared to the composite and SF regions. Higher ratios of \NII/\Ha\ have been linked to AGN activity at lower redshifts \cite{bald81}, which could indicate that there is not AGN activity. However, the lack of \NII$\lambda$6583 can be caused by several things. Nitrogen is a secondary element that is mostly created by the CNO cycle at the expense of the C and O already present in the star \citep{Chiappini2003}. The fact that nitrogen lines appear in the stacked spectra of non-broad line galaxies in the composite and SF region but not for the stacked spectra of non-broad line galaxies in the AGN region of the diagram may indicate these are galaxies with a younger star population, which has yet to start producing significant nitrogen to ionize. As stated previously, higher \CIII\ emission has been linked to metal-poor stars, making this a valid possibility as well.
This would additionally explain the other oxygen lines we see in the stacked non-broad-line galaxies in the AGN region as oxygen is a primary element created in star formation.

Further inspection of the stacked spectra does not reveal other emission lines that we would expect from an AGN. In all of the spectra we do not see \NeIV$\lambda$2423, \MgII$\lambda$2795+2802, or \NeV$\lambda$3345,3426 emission \citep{Lefevre2019,Feltre2016,Cleri2023a}. Looking at other ionization sources such as Wolf-Rayet stars, the stacked spectra do not have \NIV$\lambda$1718 or a broad \HeII$\lambda$1640 component above 5\AA\ \cite{Hainich2015, Schaerer2003, Saxena2020}. 
Other sources of high ionization could come from very massive stars, which have been linked to strong \HeII$\lambda$1640 having an equivalent width of $4-8$\AA\ for young ages or $2.5-4$\AA\ for a constant star formation rate (SFR) and extreme enrichment of nitrogen due to stellar winds \citep{Schaerer2024,Vink2023}. Due to the inability to measure the \HeII$\lambda$1640 line even in the stacked spectra, we cannot fully explore this possibility. However we note that the stacked spectrum of non-broad-line galaxies in the AGN region of the $\OIIIHg$ vs $\NeOII$ diagram has less nitrogen emission compared to the other stacked spectra.

The high-redshift BLAGN land in the expected $\OIIIHg$ vs $\NeOII$ region defined by low-redshift observations, and appear to prefer higher ratios of $\NeOII$. While investigating the non-broad line galaxies that land in the AGN region of the $\OIIIHg$ vs $\NeOII$ diagram, we conclude that these galaxies do have higher ionization than those which land in the composite region. While \NeOII\ and higher \CIII\ equivalent width require a higher ionizing potential, it is not abundantly clear what the source of this higher ionization is for these non-broad-line galaxies in the AGN region. Deeper observations of the UV region of high redshift spectra may be the key to finding a clear source or sources of ionization.

\section{Physical conditions at $z>4$} \label{ISM}

We next investigate the electron temperature $T_{e}$ and gas-phase metallicity of $z>4$ galaxies laying in each of the three regions in the $\OIIIHg$ vs $\NeOII$ diagram.

Temperature is derived using Equation (4) of \cite{Nicholls2020}:

\begin{equation}
    \log_{10}(T_{e})=\frac{3.5363+7.2939x}{1.0000+1.6298x-0.1221x^{2}-0.0074x^{3}}
\end{equation}

\noindent where $x = \log_{10}(\lambda4363/\lambda(4960+5007))$. As we cannot directly compare our measured $\OIII\lambda4363$ and $\OIII\lambda(4960+5007)$ line fluxes as they are not dust corrected, we follow the same method used in \cite{Trump22}. In \cite{Trump22}, they use the ratios $\OIII\lambda4363/\Hg$ and $\OIII\lambda(4960 + 5007)/\Hb$ along with the intrinsic Balmer ratio $\Hb/\Hg=2.1$. This results in Equation (1) from \cite{Trump22}:

\begin{equation}
    \frac{\lambda4363}{\lambda(4960+5008)}=\frac{\lambda4363}{\Hg}\left(\frac{\lambda4960+\lambda5008}{\Hb} \right)^{-1} \times (2.1)^{-1}
\end{equation}

\begin{figure}[t!]
\centering
\epsscale{1.}
\plotone{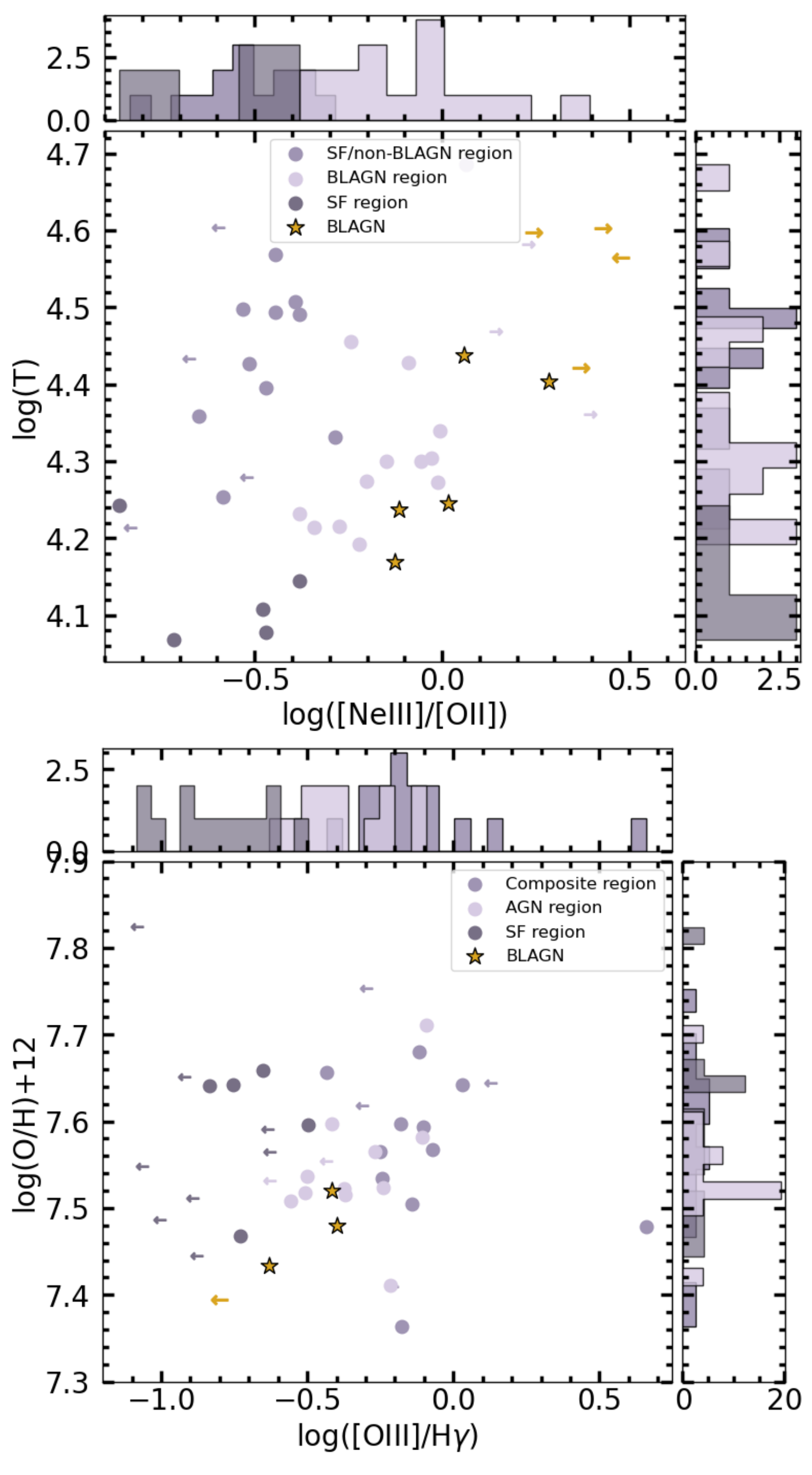}
\caption{Upper: Electron temperature of $z>4$ galaxies vs. $\NeOII$. At constant temperatures BLAGN have higher \NeOII\ than the rest of the sample. Bottom: Gas-phase metallicity of $z>4$ galaxies vs. $\OIIIHg$. There does note appear to be a trend between the two.
In both panels the colors represent the region of the $\OIIIHg$ vs $\NeOII$ diagram the galaxy lands in. From this we see that non-broad-line galaxies in the AGN region do not appear to have a different electron temperature or metallicity compared to those which land in the composite or star forming region.
\label{fig:Phys_con}} 
\end{figure}

The distribution of electron temperature and \NeOII\ ratio are shown in the top panel of Figure \ref{fig:Phys_con}. Temperatures were only calculated for galaxies that have a SNR$>3$ for \OIII$\lambda$5007,4960 \Hb, and \Hg. The top panel of Figure \ref{fig:Phys_con} shows that BLAGN have higher \NeOII\ ratios at fixed electron temperatures compared to galaxies without a broad-line feature. An Anderson-Darling test between non-broad-line galaxies occupying the AGN and composite regions returned a p-value of 0.09 for temperature. 
When we compare BLAGN to non-broad line galaxies with the same low electron temperatures, AGN have higher $\NeOII$.

Metallicity is derived for galaxies that have a 3$\sigma$ detection of $\OIII\lambda5007$ and $\OII\lambda3728$ by combining the abundances of the two previously mentioned oxygen lines by using \texttt{pyneb} \citep{Luridiana2015}. The bottom panel of Figure \ref{fig:Phys_con} plots the metallicity of these galaxies against $\OIIIHg$.
This Figure shows that BLAGN and non-broad line galaxies appear to have similar metallicities. An Anderson-Darling test between non-broad-line galaxies occupying the AGN and composite regions of the $\OIIIHg$ vs $\NeOII$ diagram returned a p-value of 0.21 for the galaxies metallicity distribution, further indicating their parent distributions are not different.  

BLAGN and non-BL galaxies have consistent distributions of temperature and (oxygen) metallicity. Similarly, galaxies in different regions of the $\OIIIHg$ vs $\NeOII$ diagram do not have significant differences in their ISM temperatures and metallicities. This indicates that the differences in line ratios between AGN and non-AGN are driven by other physical parameters, like ionization.

\section{Summary} \label{Summary}

In this work, we studied the stacked spectra of BLAGN and measured the strength of the \OIII$\lambda4363$, $\Hg$, \CIII$\lambda1908$ and \HeII$\lambda1640$ lines. From this we explore the $\OIIIHg$ vs $\NeOII$ diagram, and defined  lines to distinguish AGN and SF galaxies using 1,869 galaxies at $z\sim0$ from SDSS. We then corrected these SF/AGN lines by using the empirical evolution of the $\OIIIHg$ and \NeOII\ ratios between $z\sim0$ and $z\sim4$. These corrected lines were then compared to 90 galaxies at $z>4$ from the CEERS, JADES, and GLASS data sets. 

We studied the $\OIIIHg$ vs $\NeOII$ diagram and summarize our results as follows:
\begin{itemize}

\item Using the VO87 diagram to note AGN at $z\sim0$ Figure \ref{fig:OHNOvg_z0_final} shows that the $\OIIIHg$ vs $\NeOII$ diagram is able to separate SF galaxies from AGN. An Anderson-Darling test shows that AGN have a different parent distribution for both the \NeOII\ and $\OIIIHg$ emission line ratios, with AGN preferring higher \NeOII\ and $\OIIIHg$ ratios than the SF galaxies. 

\item There is a 0.16 dex increase in $\OIIIHg$ and 0.50 dex increase in \NeOII\ between redshifts 0 and 4. Even after accounting for this redshift evolution, the $\OIIIHg$ vs $\NeOII$ diagram is less effective for seperating BLAGN and non-BLAGN galaxies at high redshift. Although BLAGN occupy a distinct parameter space in these line ratios, non-BLAGN galaxies overlap with the same parameter so the populations cannot be reliable distinguished.

\item At $z>4$, an Anderson-Darling test between the BLAGN and rest of the sample indicates that the two populations are inconsistent with being from the same parent distribution in \NeOII\. However, the samples have similar distributions of $\OIIIHg$. This indicates the the $\OIIIHg$ vs $\NeOII$ diagram is not as effective at high redshifts, even though the AGN region of the diagram still holds the highest fraction of BLAGN (78.6\%).

\item At $z\sim0$ the SF galaxies are best modeled by Cloudy models with $-3<log(U)<-2.5$ and $0.1<Z/Z_\odot<0.2$. AGN at $z\sim0$ are best modeled with moderate ionization and moderate to high metallicity. At $z>4$ our galaxies are best represented by AGN models regardless of whether they have a broad-line feature or not. However, BLAGN are better represented by higher ionization AGN models compared to galaxies without broad emission lines.

\item Comparing stacked spectra of non-broad-line galaxies in the three regions of the $\OIIIHg$ vs $\NeOII$ diagram, we did not identify any other emission-line features that gave clear indication of potential ionization sources. Though we do note that non-BLAGN galaxies in the BLAGN region have significantly more \CIII\ emission and a lack of \NII\ emission. 

\item At constant electron temperatures, BLAGN appear to have higher values of $\NeOII$, indicating harder ionizing radiation.

\end{itemize}

Though we have enough BLAGN to do a preliminary investigation into the $\OIIIHg$ vs $\NeOII$ diagram, more insight on narrow-line AGN features is needed to test the diagnostic's ability to separate high ionization sources. This can be achieved by deeper observation that allow for stronger conclusions about the ionization sources for the non-BLAGN galaxies, allowing us to confirm or deny the presence of other AGN features such as UV line fluxes and equivalent widths. 

\section{Acknowledgements}

We acknowledge the work of our colleagues in the CEERS collaboration and everyone involved in the JWST mission.
B.E.B. and J.R.T. acknowledge support from from NASA grants JWST-ERS-01345, JWST-AR-01721, and NSF grant CAREER-1945546.
B.E.B. and A.K. acknowledge support from grants JWST-GO-03794.001. 
N.J.C. acknowledges support from NASA grants HST-AR-16609, JWST-GO-04233, and JWST-AR-05558.
Some of the data products presented herein were retrieved from the Dawn JWST Archive (DJA). DJA is an initiative of the Cosmic Dawn Center (DAWN), which is funded by the Danish National Research Foundation under grant DNRF140.


\software{\texttt{AstroPy} \citep{astropy2013}, \texttt{Matplotlib} \citep{hunter2007}, \texttt{NumPy} \citep{harris2020array}, \texttt{SciPy} \citep{jones2001},  \texttt{Linmix}}
\cite{kell07}, \texttt{eazy-py} \citep{bram08}, \texttt{pyneb} \citep{Luridiana2015}

\bibliography{lib}{}

\end{document}